\documentclass[aps,prc,twocolumn,showpacs,superscriptaddress]{revtex4}  
\usepackage{amssymb,amsmath,mathtools,color}
\usepackage{graphicx}
\usepackage{color}

\begin{document}
\title{Extrapolation of scattering data to the negative-energy region II}

\author{L. D. Blokhintsev}
\affiliation{Skobeltsyn Institute of Nuclear Physics, Lomonosov Moscow State University, Moscow 119991, Russia}
\author{A. S. Kadyrov}
\affiliation{Curtin Institute for Computation and Department of Physics and Astronomy, Curtin University, GPO Box U1987, Perth, WA 6845, Australia}
\author{A. M. Mukhamedzhanov}
\affiliation{Cyclotron Institute, Texas A\&M University, College Station, Texas 77843, USA}
\author{D. A. Savin} 
\affiliation{Skobeltsyn Institute of Nuclear Physics, Lomonosov Moscow State University, Moscow 119991, Russia}

\begin{abstract}
A problem of analytical continuation of scattering data to the negative-energy region to obtain information about bound states is discussed within an exactly solvable  potential model. This work is continuation of the previous one by the same authors [L. D. Blokhintsev et al., Phys. Rev. C \textbf{95}, 044618 (2017)]. The  goal of  this paper is to determine the most effective way of analytic continuation for different systems.  The $d+\alpha$ and $\alpha +^{12}$C systems are considered and, for comparison, an effective-range function approach and a recently suggested
$\Delta$-method [O. L. Ram\'{\i}rez Su\'arez and J.-M. Sparenberg, Phys. Rev. C {\bf 96}, 034601 (2017)] are applied. We conclude that the $\Delta$-method
is more effective for heavier systems with large values of the Coulomb parameter, whereas for light systems  with small values of the Coulomb parameter the effective-range function method might be preferable.
\end{abstract}


\maketitle

\date{Today}

\section{Introduction}

 Asymptotic normalization coefficients (ANCs) are fundamental nuclear characteristics important both in nuclear reaction and nuclear structure physics. They determine amplitudes of the asymptotic forms of bound-state nuclear wave functions in binary channels. The ANC for a virtual 
$a\leftrightarrow b+c$ process  is related directly to the residue  of the elastic $b+c$ scattering amplitude at the pole in the energy plane corresponding to the bound state of nucleus $a$ (see, e.g. Ref.~\cite{BBD}).

 The ANCs naturally appear in the expressions for the cross sections of nuclear reactions between charged particles at low energies when, due to the Coulomb barrier, the reactions occur at large distances between colliding nuclei. Astrophysical nuclear reactions represent the most important type of such reactions. The role of the ANCs in nuclear astrophysics was first discussed in 
Refs.~\cite{MukhTim,Xu}, where it was emphasized that the ANC determines the overall normalization of peripheral radiative capture reactions (see also Refs.~\cite{MukhTr,Mukh2}). The ANC method provides a powerful indirect technique in nuclear astrophysics.  

There are different ways to determine the ANCs from experimental data. 
From the peripheral reactions the ANCs can be extracted directly by normalizing the calculated cross sections to the experimental data.
However, it is impossible to directly determine the ANCs from elastic scattering data, which are measured at positive energies while the ANCs are related to the residues of the poles of the bound states at negative energies. 
Nevertheless, there is an indirect way to determine the ANC from experiment: the ANC $C_{a\to bc}$ can be determined from experimental data by extrapolating, in the plane of the center-of-mass (c.m.) energy $E$,  the partial-wave amplitude of the elastic $b+c$ scattering, obtained by the phase-shift analysis, to the pole corresponding to the bound state $a$ and lying at $E<0$. The conventional procedure for such an extrapolation is the analytic approximation of the experimental values of the  effective-range function 
(ERF) $K_l(E)$ with the subsequent continuation to the pole  (here $l$ is the orbital angular momentum). The ERF method has been successfully employed to determine the ANCs for bound (as well as resonant) nuclear states in a number of works (see, e.g. Refs.~\cite{BKSSK,SpCaBa,IrOr} and references therein). 

The ERF is expressed in terms of scattering phase shifts. In case of charged particles, 
the ERF for the short-range interaction should be modified. Such modification 
generates additional terms in the ERF. These terms depend only on the Coulomb interaction and may far exceed, in the absolute value, the informative part of the ERF containing the phase shifts. This fact hampers the practical procedure of the analytic continuation and affects its accuracy. In Ref.~\cite{Sparen} it was suggested to use for the analytic continuation the quantity $\Delta_l(E)$ [which is defined below in Section 2] rather than the ERF $K_l(E)$. Quantity $\Delta_l(E)$, which we will call a $\Delta$ function, does not contain the pure Coulomb terms. However, the validity of employing 
$\Delta_l(E)$ was not obvious, and this resulted in some discussions. It was demonstrated in Ref. \cite{BKMS} that the  $\Delta_l(E)$ function suggested in Ref.~\cite{Sparen} can be smoothly continued from the positive to the negative energy region along the real $E$ axis (see also Ref.~\cite{OrlIrg}). In what follows, using the $\Delta_l(E)$ function for extrapolation to the negative-energy region to find the ANC is referred to as a $\Delta$-method.   

The present work can be considered as a natural development and extension of Ref. \cite{BKMS} by the same authors. Here we calculate the
 scattering phase shifts and the functions $K_l(E)$ and $\Delta_l(E)$ using an analytic solution of the Schr\"odinger equation at $E>0$ with an adopted potential in the form of the square-well plus the Coulomb interaction. To the authors' knowledge, the square-well potential is the only local potential which, with the added Coulomb interaction, permits the analytic solution of the Schr\"odinger equation at any value of the orbital angular momentum.  In this approach our results are vigorous and obtained without any approximation. The calculated functions 
$K_l(E)$ and $\Delta_l(E)$ are approximated by polynomials in $E$ and extrapolated to the negative energy region including the bound-state poles of the system under consideration. This procedure imitates the approach to determining ANCs by the analytic approximation of experimental scattering data. The approximated values of $K_l(E)$, $\Delta_l(E)$, and the resulting ANCs are compared to the exact values following from the exact solution of the Schr\"odinger equation. This comparison allows one to evaluate the quality of the approximation and to compare the effectiveness of the ERF and $\Delta$-methods. 
 
Note that the simplicity of our potential model is justified by the fact that at very low energies, which we are interested in, the wave length (the reciprocal of the relative momentum of the interacting nuclei) becomes much larger than the radius of the nuclear interaction potential making  the results insensitive to the specific shape of the used potential, whether it is Woods-Saxon, square-well, delta function or anything else. 

In the present paper, the procedure described above is applied to two different nuclear systems: the $d+\alpha$ system and the $\alpha +^{12}$C system. These systems differ in the value of the Coulomb (Sommerfeld) parameter which is much larger for the latter. One more qualitative distinction between these systems is that the $d+\alpha$ system has only one bound state corresponding to the ground state of 
$^6$Li whereas the $\alpha +^{12}$C system possesses two bound states in the $0^+$ channel. One of the main results of the present paper is the conclusion that the $\Delta$-method is more effective for heavier systems with large values of the Coulomb parameter whereas for light systems  with small values of the Coulomb parameter the ERF method might be preferable.

The paper is organized as follows. 
Section II provides a brief outline 
{of} the general formalism of the elastic scattering for the superposition of a short-range and the Coulomb interactions which is necessary for the subsequent discussion. Sections III and IV deal with the $d+\alpha$ and $\alpha +^{12}$C systems, respectively. The problem of the convergence of the approximate expressions for the $\Delta$ function is discussed in Sec.~V and in the Appendix.  

Throughout the paper we use the system of units in which $\hbar=c=1$.

\section{Basic formalism}

In this section we recapitulate basic formulas which are necessary for the subsequent discussion. The formalism has been published in more detail in Ref. \cite{BKMS}.

The Coulomb-nuclear  amplitude of elastic scattering of particles $b$ and $c$ is of the form
\begin{equation}\label{fNC}
f_{NC}({\rm {\bf  k}})=\sum_{l=0}^\infty(2l+1)\exp(2i\sigma_l)\frac{\exp(2i\delta_l)-1}{2ik}P_l(\cos\theta).
\end{equation}
Here ${\rm {\bf k}}$ is the relative momentum of  $b$ and $c$, $\theta$ is the c.m. scattering angle,   
$\sigma_l=\arg\,\Gamma(l+1+i\eta)$  
 and $\delta_l$ are the pure Coulomb and Coulomb-nuclear phase shifts, respectively, and $\Gamma(z)$ is the Gamma function.
\begin{equation}\label{eta}
\eta =Z_bZ_ce^2\mu/k
\end{equation}
is the Coulomb  parameter for the $b+c$ scattering state with the relative momentum $k$ related to the energy by  $k=\sqrt{2\mu E}$,
$\mu=m_bm_c/(m_b+m_c)$, $m_i$ and $Z_ie$  are the mass and the electric charge of particle $i$.

The behavior of the Coulomb-nuclear partial-wave amplitude $f_l=(\exp(2i\delta_l)-1)/2ik$ is irregular near 
$E=0$. Therefore, one has to introduce the renormalized Coulomb-nuclear partial-wave amplitude $\tilde f_l$ \cite{Hamilton,BMS,Konig}
\begin{equation}\label{renorm}
\tilde f_l=\exp(2i\sigma_l)\,\frac{\exp(2i\delta_l)-1}{2ik}\,\left[\frac{l!}{\Gamma(l+1+i\eta)}\right]^2e^{\pi\eta}.
\end{equation}
Eq.~(\ref{renorm}) can be rewritten as 
\begin{equation}\label{renorm1}
\tilde f_l=\frac{\exp(2i\delta_l)-1}{2ik}C_l^{-2}(\eta),
\end{equation}
where $C_l(\eta)$ is the Coulomb penetration factor (or Gamow factor) determined by
\begin{align}\label{C}
C_l(\eta)&=\left[\frac{2\pi\eta}{\exp(2\pi\eta)-1}v_l(\eta)\right]^{1/2}, \\ 
v_l(\eta)&=\prod_{n=1}^{l}(1+\eta^2/n^2)\;(l>0),\quad v_0(\eta)=1.
\end{align}
It was shown in Ref. \cite{Hamilton} that  the
analytic properties of ${\tilde f}_{l}$ on the physical sheet of $E$  are analogous to the ones of the partial-wave scattering amplitude for the short-range potential and it can be analytically continued into the negative energy region.

The amplitude $\tilde f_l$ can be expressed in terms of the Coulomb-modified ERF $K_l(E)$ \cite{Hamilton, Konig}  by
\begin{align} 
\label{fK}
\tilde f_l&=\frac{k^{2l}}{K_l(E)-2\eta k^{2l+1}h(\eta)v_l(\eta)}\\ 
&=\frac{1}{kC_l^2(\eta)(\cot\delta_l-i)} \\ 
&=\frac{1}{v_l^2\Delta_l(E)-ikC_l^2(\eta)},
\label{fK3}
\end{align}   
where
\begin{align}\label{scatfun}
K_l(E)&= k^{2l+1} \left[ C_l^2(\eta)(\cot\delta_l-i) + 2 \eta h(k)v_l(\eta) \right],\\ 
h(\eta) &= \psi(i\eta) + \frac{1}{2i\eta}-\ln(i\eta), \\  
\Delta_l(E)&=kC_0^2(\eta)\cot\delta_l, 
\label{Deltal}
\end{align}
and $\psi(x)$ is the digamma function. $\Delta_l(E)$ is the $\Delta$ function introduced in \cite{Sparen}. 

It was shown in \cite{Hamilton} that function $K_l(E)$ defined by (\ref{scatfun}) is analytic near $E=0$ and can be expanded into Taylor series in $E$. In the absence of the Coulomb interaction ($\eta=0$) $K_l(E)=k^{2l+1}\cot\delta_l(k)$.

If the $b+c$ system has in the partial wave $l$ the bound state $a$ with the binding energy $\varepsilon=\varkappa^2/2\mu>0$, then the amplitude $\tilde f_l$ has a pole at $E=-\varepsilon$. The residue of $\tilde f_l$ at this point is expressed in terms of the ANC
$C^{(l)}_{a\to bc}$ \cite{BMS} as
\begin{align}\label{res2}
{\rm res}\tilde f_l(E)|_{E=-\varepsilon}&=\lim_{\substack{E\to -\varepsilon}}[(E+\varepsilon)\tilde f_l(E)] \\
&=
-\frac{1}{2\mu}\left[\frac{l!}{\Gamma(l+1+\eta_b)}\right]^2 \left[C^{(l)}_{a\to bc}\right]^2,
\label{res22}
\end{align}
where $\eta_b=Z_bZ_ce^2\mu/\varkappa$ is the Coulomb  parameter for the $b+c$ bound state $a$.

In what follows, the short-range nuclear interaction between particles $b$ and $c$ is described by the square well potential
\begin{align}\label{potential}
V(r)=\left\{ \begin{matrix} 
 -V_0 & {\rm if  } & 0\le r\le R\\
 0 & {\rm if  } & r > R 
\end{matrix} \right. ,
\end{align}
where $R$ is the radius of the square well and $V_0>0$ is its depth.

The solution of the Schr\"odinger equation for the potential (\ref{potential}) plus the Coulomb interaction results in the following expression for the phase shift $\delta_l$ \cite{BKMS}
\begin{align}
\label{cotdelta}
\cot\delta_l  & \nonumber \\
=&\dfrac{\dfrac{d\hat G_{l,\eta}(k,R)}{dR} \hat F_{l,\eta_1}(K,R)
- \dfrac{d\hat F_{l,\eta_1}(K,R)}{dR} \hat G_{l,\eta}(k,R)} 
{\dfrac{d\hat F_{l,\eta}(k,R)}{dR} \hat F_{l,\eta_1}(K,R)
- \dfrac{d\hat F_{l,\eta_1}(K,R)}{dR} \hat F_{l,\eta}(k,R)} .
\end{align}
Here $K=\sqrt{2\mu(E+V_0)}$, $\hat F_{l,\eta}(q,r)= F_l(\eta, qr)/qr$, $\hat G_{l,\eta}(q,r)= -G_l(\eta, qr)/qr$, $F_l(\eta, \rho)$ and 
$G_l(\eta, \rho)$ are the regular and irregular Coulomb functions, respectively~\cite{DLMF}.

Eq.(\ref{cotdelta}) allows one to calculate the functions $K_l(E)$ and $\Delta_l(E)$ using Eqs. (\ref{scatfun}) and (12).
Detailed derivation and explicit analytic expressions for $K_l(E)$ and $\Delta_l(E)$ are given in \cite{BKMS}.

\section{$d+\alpha$ system}

Consider the $d+\alpha$ system having one bound state corresponding to the ground state of $^6$Li with $l=0$. For this system 
$m_b=m_d$=1877.79 MeV, $m_c=m_\alpha$=3727.379 MeV, $m_a=m_{^6\mathrm{Li}}$=5601.518 MeV, $Z_bZ_c$=2, binding energy 
$\varepsilon=m_d+m_\alpha-m_{^6\mathrm{Li}}$=1.474 MeV. 

Parameters of the square well $V_0$=7.400955728 MeV and $R$=3.963659401 fm were found by fitting the binding energy and the ANC $C^{(0)}_{^{6}{\rm Li}\to \alpha d}$=2.29 fm$^{-1/2}$ obtained in Ref. \cite{BKSSK}.  
For brevity ANC $C^{(0)}_{^{6}{\rm Li}\to \alpha d}$ will be denoted as $C$.

\subsection{Approximation of the ERF for the $d+\alpha$ system by the Taylor series}

Consider first the approximation of the ERF  $K_0(E)$ by the Taylor series in $E$ at $E=0$. Expansion into the Taylor series is performed using analytic expressions (\ref{scatfun}) and (\ref{cotdelta}). In fact we limit ourselves by the first several terms of the expansion. A polynomial obtained this way is then continued analytically to the negative-energy region to the bound-state pole. 

Two versions of the approximation are considered:
\begin{itemize}
\item []
Version~1. Both the binding energy and the ANC are found from the approximated form of $K_0(E)$.
\item []
Version~2. The binding energy is preset ($\varepsilon$=1.474 MeV) and only the ANC is sought. 
\end{itemize}
Actually, in the second version we approximate the function 
$F(E)=(K_0(E)-K_b)/(E+\varepsilon)$, where $K_b=2\eta k h(\eta)|_{E=-\varepsilon}$ is the value of $K_0(E)$ at $E=-\varepsilon$. Function 
$F(E)$ is finite at $E=-\varepsilon$ and its approximation by the Taylor series guarantees the correct value of $K_0(E)$ at 
$E=-\varepsilon$, which is the correct position of the pole of the scattering amplitude corresponding to the bound state.    
		
The results of the calculation of the binding energy (in the first version) and the ANC are presented in Table~\ref{table1}.  In this table, as well in all the following tables, $N$ denotes the power of the approximating polynomial. The exact values of the corresponding quanities obtained by the exact calculations within the model used are shown in the last line of Table \ref{table1}.	 One can see that the convergence in 
$N$ is quite good, especially within the second version.

\begin{table}[htb]
\caption{Approximation of $K_0(E)$} for the $d+\alpha$ system.  
\begin{center}
\begin{tabular}{|c|c|c|c|c|}
\hline
 & \multicolumn{2}{c|}{Version 1} & \multicolumn{2}{c|}{Version 2} \\  
\hline 
N & $\varepsilon$, MeV & $C$, fm$^{-1/2}$ & $\varepsilon$, MeV & $C$, fm$^{-1/2}$  \\ 
\hline 
2 & 1.4546 &  2.256 & 1.474 & 2.894  \\
3 & 1.4729 & 2.2858 & 1.474 & 2.2902  \\ 
4 & 1.4744 & 2.2917 & 1.474 & 2.28997 \\
\hline
exact value & 1.474 & 2.29 & 1.474 & 2.29 \\
\hline
\end{tabular}
\end{center}
\label{table1}
\end{table} 

\begin{figure}[htb]
\center{\includegraphics[width=0.8\linewidth]{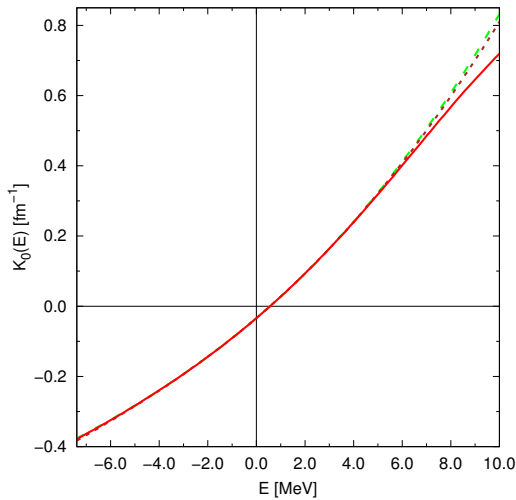}}
\caption {The $K_{0}(E)$ function for the $d+ \alpha$ system. 
The solid red line is the exact $K_{0}(E)$ function; the green dashed line is the approximation of $K_{0}(E)$ by the Taylor polynomial when the binding energy and the ANC of the bound state $(d\,\alpha)$ are not fixed (version 1);  the brown dotted line is approximation of $K_{0}$ by the Taylor polynomial of the third order when only the binding energy of $(d\,\alpha)$ is fixed (version 2).}
\label{pic1_1_2}
\end{figure}

The exact function $K_0(E)$ for the $d+\alpha$ system and its approximations by the polynomial of the third power in $E$ ($N$=3) are shown in 
Fig. \ref{pic1_1_2} for two versions of the approximation. It is seen that $N$=3 ensures a quite good description of the exact ERF $K_0(E)$ over the wide energy interval.

\subsection{Approximation of the $\Delta$ function for the $d+\alpha$ system by the Taylor series}

In this subsection we will consider the function $\mathrm{Re}[D_0(E)]=K_0(E)-\mathrm{Re}[2\eta k h(\eta)]$ which is the real part of the denominator $D_0(E)$ of the partial-wave amplitude $\tilde f_0(E)$ for the $d+\alpha$ system. At $E<0$ $\mathrm{Re}[D_0(E)]=
\tilde f_0^{-1}(E)$ and the condition $\mathrm{Re}[D_0(E)]=0$ is the condition of a pole of $\tilde f_0(E)$ corresponding to the bound state. 
At $l=0$ $\mathrm{Re}[D_0(E)]$ coincides with the function $\Delta_l(E)$ (see Eq. (\ref{Deltal})) introduced in Ref.~\cite{Sparen}. Therefore, in what follows we will use the notation
$\Delta_0(E)$ instead of $\mathrm{Re}[D_0(E)]$. 

As in the case of $K_0(E)$ (see Section 3A), we will approximate  $\Delta_0(E)$ by the Taylor series in $E$ at $E=0$ with the subsequent continuation to the negative-energy region. We consider the same two versions of the approximation as in Section 3A, however, in the first  
{version} we now use $\Delta_0(E)$ rather than $K_0(E)$. In addition, in the second  
{version} we actually approximate the function 
$\Delta_0(E)/(E+\varepsilon)$.

The results of the approximation of $\Delta_0(E)$ by the first several terms of the Taylor series are presented in Table~\ref{table3}. A dash means that the given approximation does not lead to the bound state.  The result marked by an asterisk is related to the fact that in the $N=3$ approximation the function $\Delta_0(E)$ turns into zero to the right of the point $E$=-1.474 MeV.

\begin{table}[htb]
\caption{Approximation of $\Delta_0(E)$} for the $d-\alpha$ system.  
\begin{center}
\begin{tabular}{|c|c|c|c|c|}
\hline
 & \multicolumn{2}{c|}{Version 1} & \multicolumn{2}{c|}{Version 2} \\  
\hline 
N & $\varepsilon$, MeV & $C$, fm$^{-1/2}$ & $\varepsilon$, MeV & $C$, fm$^{-1/2}$  \\ 
\hline 
2 & -      & -        & 1.474      & 0.799  \\
3 & 0.432  & 0.565    & 0.493$^*$  & 0.669  \\
4 & -      & -        & 1.474      & 0.087  \\
\hline
exact values & 1.474 & 2.29  & 1.474      & 2.29  \\
\hline
\end{tabular}
\end{center}
\label{table3}
\end{table} 

\begin{figure}[htb]
\center{\includegraphics[width=0.8\linewidth]{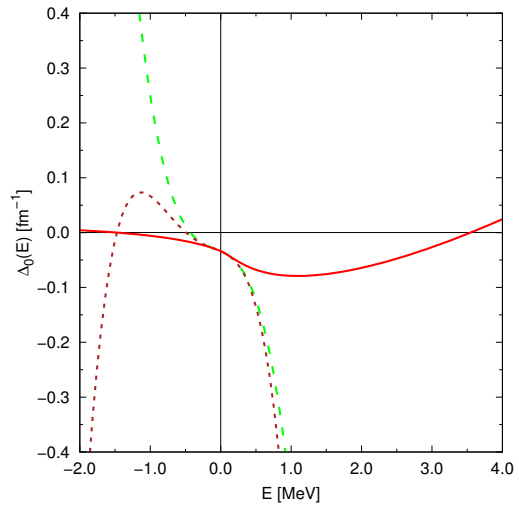}}
\caption {The $\Delta_0(E)$ function for the $d+ \alpha$ system. The notations are the same as in Fig. \ref{pic1_1_2} but for $\Delta_0(E)$.}
\label{pic2_1_2}
\end{figure}

The exact function $\Delta_0(E)$ for the $d+\alpha$ system and its approximations by the polynomial of the third power in $E$ ($N$=3) are shown in Fig.~\ref{pic2_1_2} for two versions of the approximation.
It is seen from Table~\ref{table3} and Fig.~\ref{pic2_1_2} that the employed approximation of $\Delta_0(E)$ is absolutely unsatisfatory.

\section{$\alpha+^{12}$C system with two bound $0^+$ states}

The goal of this paper is to find out which of the two extrapolation methods, the Coulomb-modified 
ERF $K_{0}(E)$ or the 
{Ram\'{\i}rez Su\'arez-}Sparenberg function $\Delta_{0}(E)$ \cite{Sparen}, works better for the  $\alpha + {}^{12}{\rm C}$ system in the  $l=0$ partial wave with the ground and excited $0^{+}$ bound states. To determine it we use 
{the same simple model as for the $d+\alpha$ system, namely} a square-well nuclear potential plus the Coulomb interaction acting between two point-like particles $\alpha$ and ${}^{12}{\rm C}$. In the realistic potential approach the wave function of the relative motion of $\alpha-{}^{12}{\rm C}$ has two nodes at $r>0$ for the ground bound state. In our simplified model we use one potential supporting two $0^{+}$ bound states, the ground and the first excited ones. In this simplified approach the ground bound-state 
wave function of the $\alpha-{}^{12}{\rm C}$ system is nodeless at $r>0$ while the wave function of the first excited state has one node at $r>0$. 

For the $\alpha+^{12}$C system, we have
$m_b=m_\alpha$=3727.379 MeV, $m_c=m_{^{12}\mathrm{C}}$=11174.862 MeV, $m_a=m_{^{16}\mathrm{O}}$ =14895.079 MeV, $Z_b Z_c$=12.
   
We adopt the square well potential with parameters $V_0$=13.70363036 MeV and $R$=6.009708703 fm. The sum of this nuclear potential and the Coulomb interaction leads to two bound $0^+$ states wth the binding energies $\varepsilon_1= 1.113$ MeV and 
$\varepsilon_2= 7.162$ MeV. These binding energies coincide with the experimental ones. The ANC values for such a potential are $C_1$=3218.458518 fm$^{-1/2}$ and $C_2$=3475.353169 fm$^{-1/2}$ for the excited and ground states, respectively. Because we use a simplified potential model these ANCs
should not be considered as realistic ones but they will help us to identify which extrapolation method works better for the $\alpha + {}^{12}{\rm C}$ system. 

Note that in principle one may use an alternative way to find the parameters $V_0$ and $R$, namely, by fitting them to the  value of 
$\varepsilon_1$ and to the value of $C_1$ obtained from the analysis of experimental data, e.g. from Ref. \cite{Avila}. The qualitative results stated below do not depend on the way how the square-well parameters are chosen.

Since the considered $\alpha+^{12}$C system has two bound states, the ERF $K_0(E)$, as well as function $\Delta_0(E)$,  
has two poles: one at negative energy ($E_{i2}$) and another at positive energy ($E_{i1}$). The pole at negative energy leads to the change of the sign of the partial-wave amplitude $\tilde f_0(E)$ in the interval between the points corresponding to the two bound states. This guarantees the correct signs of the residues of $\tilde f_0$ at both poles $E=-\varepsilon_1$ and $E=-\varepsilon_2$ (see Refs.~\cite{BlSav,Baz}). 
 As is seen from Eq.~(\ref{res22}), the sign of both residues should be negative in order to guarantee that the ANC is real. The pole at $E>0$ is due to the Levinson theorem. The above mentioned values of $V_0$ and $R$ result in $E_{i2}=-4.48135$ MeV and $E_{i1}$=25.315 MeV.

\subsection{Approximation of the ERF for the  $\alpha+^{12}$C system: search for the parameters of the excited $0^+$ state}

The approximation versions 1 and 2 are similar to those for the $d+\alpha$ system. Within the version 2 the binding energy 
$\varepsilon_1$ of the excited state is fixed. The presence of the ground state and of the pole at $E_{i2}$ is not taken into account explicitly. 

\begin{table}[htb]
\caption{Approximation of $K_0(E)$} for the $\alpha-^{12}$C system.  
\begin{center}
\begin{tabular}{|c|c|c|c|c|}
\hline
 & \multicolumn{2}{c|}{Version 1} & \multicolumn{2}{c|}{Version 2} \\  
\hline 
N & $\varepsilon_1$, MeV& $C_1$, fm$^{-1/2}$ & $\varepsilon_1$, MeV & $C_1$, fm$^{-1/2}$  \\ 
\hline 
2 & 0.457  & 14361  & 1.113 & 10928  \\
3 & - & - & 1.113 & 3090.07  \\ 
4 & 1.042 & 3060.34  & 1.113 & 3230.43 \\
5 & 1.122 & 3265.97 & 1.113 & 3217.94 \\
6 & 1.1126 & 3215.71 & 1.113 & 3216.71 \\
\hline
exact values & 1.113 & 3218.46 & 1.113 & 3218.46 \\
\hline
\end{tabular}
\end{center}
\label{table2}
\end{table}

\begin{figure}[htb]
\center{\includegraphics[width=0.8\linewidth]{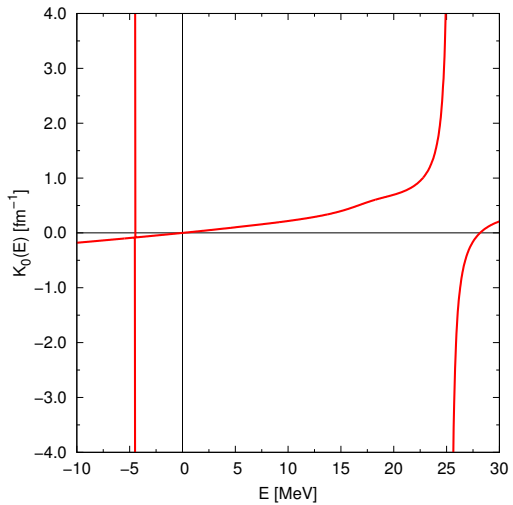}}\\
\caption {The exact $K_{0}(E)$ function for the system $\alpha + {}^{12}{\rm C}$ with two $0^{+}$ bound states. The pole at the negative energy is very narrow.}\label{pic1_2_1}
\end{figure}

\begin{figure}[htb]
\center{\includegraphics[width=0.8\linewidth]{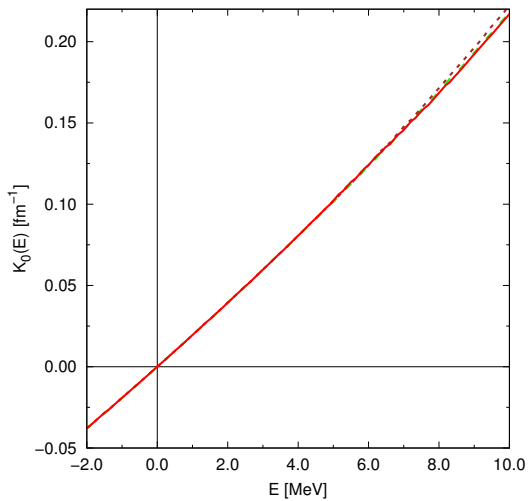}}\\
\caption {The same as in Fig. \ref{pic1_1_2} but for the system $\alpha +{}^{12}{\rm C}$.}\label{pic1_2_2}
\end{figure}

The results of the calculations are presented in Table~\ref{table2} and Figs.~\ref{pic1_2_1} and \ref{pic1_2_2}.
The exact function $K_0(E)$ for the  $\alpha+^{12}$C system in the $0^+$ channel is shown in Fig.~\ref{pic1_2_1} in a wide energy interval. 
Note that $K_0(E)$ is not equal to zero at $E=0$, however it is rather small: $K_0=-1.609\, 10^{-6} \,\mathrm{fm}^{-1}$. This fact leads to a large value of the scattering length.  
In Fig.~\ref{pic1_2_2} we present the exact function $K_0(E)$ and its approximation by the polynomial of the third power in $E$ for two versions of the approximation. The energy interval is much more narrow than in Fig.~\ref{pic1_2_1} and does not include the poles of $K_0(E)$ which cannot be described by the polynomial approximation.    
It is seen from Table~\ref{table2} that, although the results are quite satisfactory, the convergence to the exact values is slower than in the case of the $d+\alpha$ system.  

\begin{widetext}
\begin{table}[htb]
\center
\caption{Approximation of $\Delta_0(E)$} for the $\alpha+^{12}$C system.  
\begin{center}
\begin{tabular}{|c|c|c|c|c|c|c|}
\hline
 & \multicolumn{2}{c|}{Version 1} & \multicolumn{2}{c|}{Version 2} & \multicolumn{2}{c|}{Version 3} \\  
\hline 
N & $\varepsilon_1$, MeV & $C_1$, fm$^{-1/2}$ & $\varepsilon_1$, MeV & $C_1$, fm$^{-1/2}$ & $\varepsilon_1$, MeV  & $C_1$, fm$^{-1/2}$ \\ 
\hline 
2 & -     & -        & 1.113  & 2813.41  & 1.113  & 3211.95 \\
3 & 0.915 & 3296.90  & 1.113  & 3421.48  & 1.113  & 3224.88 \\
4 & -     & -        & 1.113  & 3153.03  & 1.113  & 3216.54 \\
5 & 1.064 & 3048.47  & 1.113  & 3245.73  & 1.113  & 3219.89 \\
6 & 1.147 & 3476.67  & 1.113  & 3205.76  & 1.113  & 3217.52 \\
7 & 1.100 & 3131.55  & 1.113  & 3226.06  & 1.113  & 3219.26 \\
\hline
exact values & 1.113 & 3218.46  & 1.113  & 3218.46  & 1.113  & 3218.46 \\
\hline
\end{tabular}
\end{center}
\label{table4}
\end{table} 
\end{widetext}

\subsection{Approximation of the $\Delta$ function for the $\alpha+^{12}$C system by the Taylor series: search for the parameters of the excited $0^+$ state}

Consider three versions of the approximation: 
\begin{itemize}
\item []
Version~1. Both the binding energy and the ANC are found from the approximated form of $\Delta_0(E)$.
\item []
Version~2. The binding energy is preset ($\varepsilon_{1}$=1.113 MeV) and only the ANC is sought. Function 
$\Delta_0(E)/(E+\varepsilon_1)$ is approximated.
\item []
Version~3. The binding energy is preset ($\varepsilon_{1}$=1.113 MeV) and only the ANC is sought. Function 
$\ln(\Delta_0(E)/(E+\varepsilon_1))$ is approximated by the Taylor expansion.
\end{itemize}
Using version 3 is related to the fact that the $\Delta$ function changes drastically near $E=0$. 

\begin{figure}[htb]
\center{\includegraphics[width=0.8\linewidth]{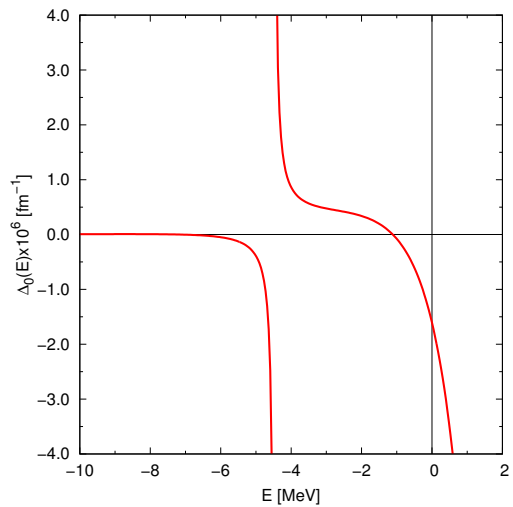}}\\
\caption {The exact $\Delta_0(E)$ function for the system $\alpha+ {}^{12}{\rm C}$ with two bound states. The pole at negative energy is located at $E_{i2}=-4.48135$ MeV.}
\label{pic2_2_1}
\end{figure}

\begin{figure}[htb]
\center{\includegraphics[width=0.8\linewidth]{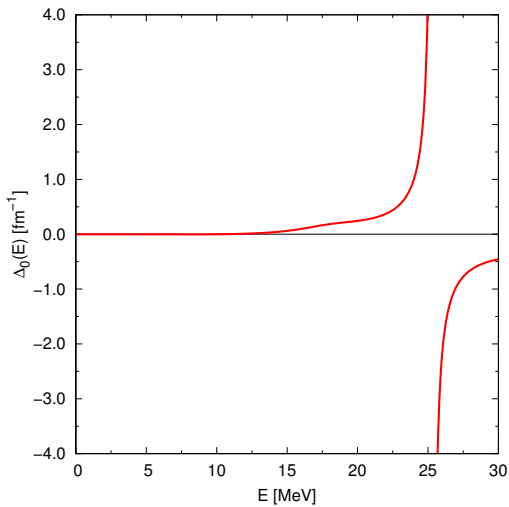}}\\
\caption {The exact $\Delta_0(E)$ function for the system $\alpha+ {}^{12}{\rm C}$ with two bound states. The pole at positive energy is located at $E_{i1}= 25.315$ MeV.}
\label{pic2_2_2}
\end{figure}

\begin{figure}[htb]
\center{\includegraphics[width=0.8\linewidth]{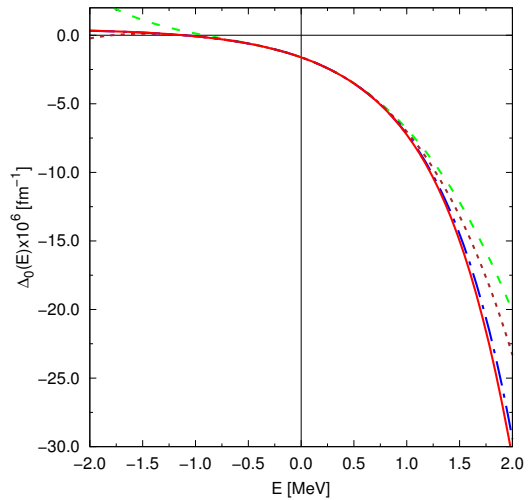}}\\
\caption {The $\Delta_0(E)$ function for the $\alpha+ {}^{12}{\rm C}$ system.
The solid red line is the exact $\Delta_0(E)$ function; the green dashed line is the approximation of the $\Delta_0(E)$ function by the Taylor polynomial of the third order when the binding energy and the ANC of the bound state $(\alpha\,{}^{12}{\rm C})$ are not fixed;  the brown dotted line is approximation of $\Delta_0(E)$ by the Taylor polynomial of the third order when the binding energy of $(\alpha\,{}^{12}{\rm C})$ is fixed while the ANC is a fitting parameter; the blue dash-dotted line is obtained using the approximation of $\ln(\Delta_0(E)/(E+\varepsilon_1))$ by the third order Taylor polynomial when the binding energy and the ANC of the bound state $(\alpha\,{}^{12}{\rm C})$ are not fixed.}
\label{pic2_2_3}
\end{figure}

The results of the calculations are presented in Table~\ref{table4}.
The exact $\Delta_0(E)$ function for the $\alpha+^{12}$C system is shown in Figs.~\ref{pic2_2_1} and \ref{pic2_2_2} for different energy intervals.
Figure~\ref{pic2_2_3} 
presents the exact  $\Delta_0(E)$ function for the $\alpha+^{12}$C system and its approximation by the polynomial of the third power in $E$ corresponding to the aforementioned three versions of the approximation.
As one can see from Table~\ref{table4} and Fig.~\ref{pic2_2_3},
the polynomial approximation of the $\Delta_0(E)$ function for the $\alpha+^{12}$C system, in contrast to the lighter  $d+\alpha$ system, turns out to be a quite good approximation.

\subsection{Approximation of the $\Delta$ function for the $\alpha+^{12}$C system by the Taylor series: search for the parameters of the ground $0^+$ state}

If one intends to determine the ANC $C_2$ for the ground $0^+$ state, it is necessary to explicitly include in the approximation form of 
$\Delta_0(E)$ the presence of the pole $E_{i2}$ at $E<0$. 

Consider two versions of the approximation:
\begin{itemize}
\item []
Version~1. Approximation of the function ${\Delta_0(E)(E-E_{i2})}/{(E+\varepsilon_1)(E+\varepsilon_2)}$
\item []
Version~2. Approximation of the function $\ln\left[-{\Delta_0(E)(E-E_{i2})}/{(E+\varepsilon_1)(E+\varepsilon_2)}\right]$.
\end{itemize}
Within both versions the positions of two bound states and of the pole $E_{i2}$ are preset. The pole $E_{i1}$ lies far from the negative energy region and its influence can be ignored.

\begin{table}[htb]
\caption{Approximation of $\Delta_0(E)$ for the $\alpha+^{12}$C system taking into account the ground $0^+$ state.}
\begin{center}
\begin{tabular}{|c|c|c|c|c|}
\hline
 & \multicolumn{2}{c|}{Version 1} & \multicolumn{2}{c|}{Version 2}  \\  
\hline 
N & $C_1$, fm$^{-1/2}$ & $C_2$, fm$^{-1/2}$ & $C_1$, fm$^{-1/2}$ & $C_2$, fm$^{-1/2}$ \\ 
\hline 
2 & 2714.48  & 72.32  & 3204.12 & 1953.26 \\
3 & 3496.00  &   -    & 3223.23 & 9520.82 \\
4 & 3132.61  &  29.62 & 3216.19 &  223.77 \\
5 & 3254.24  &   -    & 3219.81 & 5.6 10$^7$ \\
\hline
exact values & 3218.46  & 3475.35 & 3218.46 & 3475.35 \\
\hline
\end{tabular}
\end{center}
\label{table5}
\end{table}

\begin{figure}[htb]
\center{\includegraphics[width=0.8\linewidth]{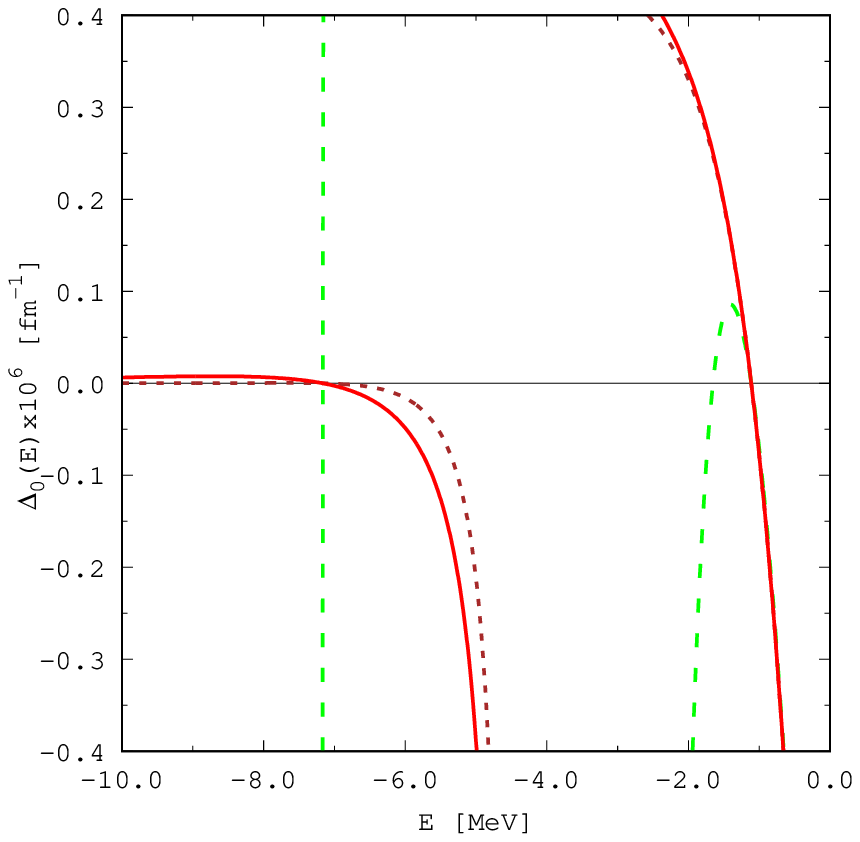}}\\
\caption {The $\Delta_0(E)$ function for the system $\alpha+ {}^{12}{\rm C}$ with two $0^{+}$ bound states.
The solid red line is the exact $\Delta_0(E)$; the brown dotted line is obtained using approximation of the function ${\Delta_0(E)(E-E_{i2})}/{(E+\varepsilon_1)(E+\varepsilon_2)}$ by the Taylor polynomial of the third order; the blue dash-dotted line corresponds to the approximation of the function $\ln\left[-{\Delta_0(E)(E-E_{i2})}/{(E+\varepsilon_1)(E+\varepsilon_2)}\right]$ by the Taylor polynomial of the third order.}
\label{pic2_3_1}
\end{figure}


As before the approximation is based on the Taylor expansion at $E=0$. 
The results of the calculations with the two versions of the approximation are presented in Table~\ref{table5} and in Fig.~\ref{pic2_3_1}.
A dash in the table means that the given version of the approximation gives a wrong sign for the derivative of $\Delta_0(E)$ at $E=-\varepsilon_2$ and, therefore, does not lead to a genuine bound state.
It is clear from Table~\ref{table5} that the approximation used here does not allow one to obtain any reasonable result for the ANC $C_2$ corresponding to the ground state of $^{16}$O even if one presets explicitly the position of the pole of $\Delta_0(E)$ at $E<0$. This is not surprising since the ground state is located far from the point $E=0$ at which the expansion in $E$ is performed. The situation gets much worse if one tries to determine $C_2$ by extrapolating the experimental data since the position of the pole $E_{i2}$ is not known from the experiments.

Note that the attempts to determine $C_2$ by extrapolating the function $K_0(E)$ or $\Delta_0(E)$ from the positive to the negative energy region were made in Refs. \cite{Orl1,OrlIrg}. In these papers, the parameters of  the analytic approximation of $K_0(E)$ and $\Delta_0(E)$ were fitted to the results of the phase-shift analysis of the elastic $\alpha-^{12}$C scattering at low energies. However, the $C _2$ values presented in these papers could hardly be taken seriously for the following reasons. In Ref. \cite{Orl1}, while continuing the $K_0(E)$ function to the point corresponding to the ground state, the authors ignored the presence of the excited $0^+$ state which affects significantly 
the behavior of $K_0(E)$ at $E<0$. In Ref. \cite{OrlIrg} the excited state was taken into account, however, the approximated analytic form of 
$\Delta_0(E)$ used by the authors ignored the existence of the pole of $\Delta_0(E)$ at $E<0$. This fact is the reason for the wrong sign of the residue of the partial-wave scattering amplitude $\tilde f_0$ at the pole corresponding to the ground state. It leads to an unphysical imaginary value of the ANC $C_2$. Furthermore, the real value of $C_2$ presented in Ref. \cite{OrlIrg} is also erroneous. This is due to the improper manipulation with the absolute value sign for the residue of $\tilde f_0$. 

It is worth mentioning that the exact partial-wave $\alpha+^{12}$C scattering amplitudes, in contrast to our theoretical model, possess a number of singularities (branching points) situated at $E<0$ between the ground and excited $0^+$ state poles. These singularities are due to the following Feynman diagrams contributing to the elastic $\alpha+^{12}$C scattering amplitude:
\begin{enumerate}
\item [1)]
The loop diagram describing two-pion exchange between $\alpha$ and $^{12}$C.
\item [2)]
The pole diagram describing the $^8$Be transfer process (or the loop diagram describing two-$\alpha$ transfer).
\item [3)]
The triangle diagrams describing scattering of an $\alpha$ particle on virtual nucleons containing in $^{12}$C. 
\end{enumerate}
These singularities should be taken into account in analytic continuation of  scattering data to the ground-state pole. It is obvious that the approximation of the $K_0(E)$ or $\Delta_{0}(E)$ function by polynomials or rational functions cannot take into account the presence of these singularities.

\section{Convergence of the approximation for the $\Delta$ function}

The renormalized Coulomb-nuclear partial-wave scattering amplitude $\tilde f_0(E)$ can be written as follows ($l=0$)
\begin{align}\label{tilde1}
\tilde f_0(E)&=1/D_0(E), 
\end{align}
where
\begin{align}
D_0(E)&=K_0(E)-R(E), \\
\label{regfun}
R(E)&= 2\alpha_1 h(\eta), \\
h(\eta)&=2\alpha_1 (\psi(i\eta)-\ln(i\eta)+1/(2i\eta)),
\end{align}
and $\alpha_1= z_b z_c e^2 \mu>0$, $\eta= \alpha_1/\sqrt{2\mu E}$. 
We remind that the $\Delta$ function for $l=0$, $\Delta_0(E)$, which we are interested in is directly related to $D_0(E)$: $\Delta_0(E)=\mathrm{Re}[D_0(E)]$.

It is known that the ERF $K_0(E)$ can be expanded in powers of $E$. In order to decide on the problem of similar power expansion and the polynomial approximation for the whole denominator $D_0(E)$ (and hence for $\Delta_0(E)$), we consider the properties of the function 
$h(\eta)$.

Since at $E \to 0$ $\eta \to \infty$, one may use the asymptotic expansion for $\psi(i\eta)$ \cite{Olver} which results in the following expansion of $h(\eta)$:
\begin{align}
h(\eta)&=- \sum_{\nu=1}^\infty\frac{B_{2\nu}}{2\nu(i\eta)^{2\nu}} \\
&= - \sum_{\nu=1}^\infty\frac{B_{2\nu}}{2\nu} \left( \frac{-2\mu E}{\alpha_1^2} 
\right)^\nu  \\
\label{h-asymptE}
&=- \sum_{\nu=1}^{n-1}\frac{B_{2\nu}}{2\nu} \left( \frac{-2\mu E}{\alpha_1^2} \right)^\nu -U_n(E), \\
&\equiv h_n(\eta)-U_n(E), 
\label{hn}
\end{align}
where $B_{2\nu}$ are the Bernoulli numbers. At $n=1$ the sum in (\ref{h-asymptE}) is equal to zero.
The form and the features of the residual term $U_n(E)$ are considered in the Appendix. In the present section we consider the separate terms of the expansion (\ref{h-asymptE}).

The series (\ref{h-asymptE}) can be considered as the expansion in $E$. However, due to the features of the Bernoulli numbers, the series 
(\ref{h-asymptE}) is asymptotic, that is, divergent. Nevertheless, it is worthwhile to investigate 
the first few terms in 
(\ref{h-asymptE}) which contribute to the polynomial approximation of $\Delta_0(E)$.  
The rate of convergence of the series (\ref{h-asymptE}) at given $E$ is determined by the quantity $\alpha_2=2 \mu/ \alpha_1^2$. The smaller $\alpha_2$, the faster is the convergence. 

\begin{figure}[htb]
\center{\includegraphics[width=0.8\linewidth]{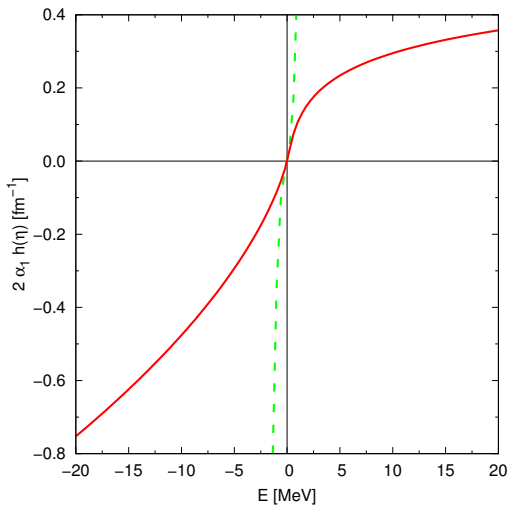}}\\
\caption {The real part $\mathrm{Re}[R(E)]$ for the $d+ \alpha$ system. The solid red line is the exact result; the dashed green line is the asymptotic expansion of $\mathrm{Re}[R(E)]$ up to $E^{3}$ including.}\label{pic3_1_2}
\end{figure}

\begin{figure}[htb]
\center{\includegraphics[width=0.8\linewidth]{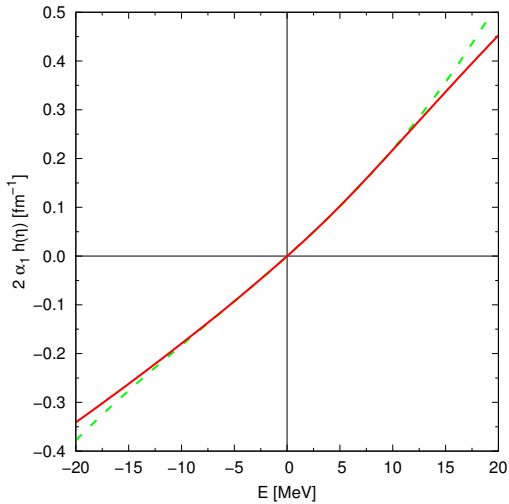}}\\
\caption {The real part $\mathrm{Re}[R(E)]$ for the $\alpha+ {}^{12}{\rm C}$ system. The solid red line is the exact result; the dashed green line is the asymptotic expansion of $\mathrm{Re}[R(E)]$ up to $E^{3}$ including.}\label{pic3_1_4}
\end{figure}

For the $d+\alpha$ system the value of $\alpha_2$ is rather large: $\alpha_2=7.53$ MeV$^{-1}$. As a result, the approximation of $h(\eta)$ by the first terms of the series (\ref{h-asymptE}) is poor. This is seen in Fig. \ref{pic3_1_2} which displays the real part $\mathrm{Re}[R(E)]$ for the $d+ \alpha$ system.
However, for the heavier $\alpha+^{12}$C system   $\alpha_2$ is two orders of magnitude smaller than for the $d+\alpha$ system: 
$\alpha_2=0.0933$ MeV$^{-1}$. Therefore, for this system $h(\eta)$ can be successfully approximated by first few terms of the expansion 
 (\ref{h-asymptE}) in a wide energy  interval. This result is illustrated in Fig.~\ref{pic3_1_4} which displays the real part $\mathrm{Re}[R(E)]$ for the $\alpha+ {}^{12}{\rm C}$ system.

\begin{figure}[htb]
\center{\includegraphics[width=0.8\linewidth]{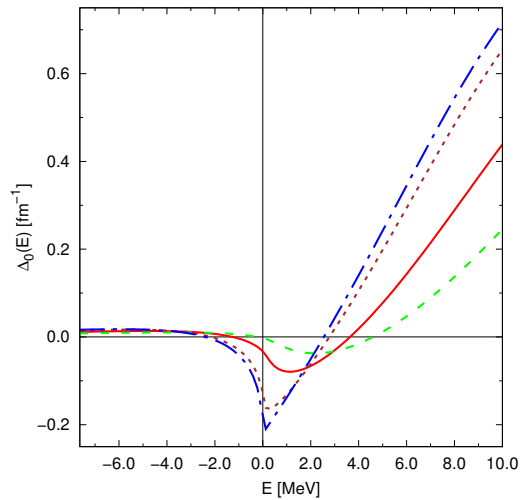}}\\
\caption {The dependence of the $\Delta_{0}(E)$ function on the charge-scaling factor $\beta$ (see the text) for $d+ \alpha$. Solid red line, $\beta =1$; dashed green line, $\beta =2$; dotted brown line, $\beta=0.2$; dash-dotted blue line, $\beta = 0.$}
\label{picture}
\end{figure}

Thus, one can conclude that the polynomial approximation of the function $R(E)=2\alpha_1h(\eta)$ and hence of the functions $D_0(E)$ and 
$\Delta_0(E)$ is more effective for systems with larger values of the product of charges $Z_bZ_ce^2$ and the reduced mass $\mu$. This inference is clearly demonstrated in Fig. \ref{picture} which displays the calculations of $\Delta_0(E)$ for the $d+\alpha$ system obtained by substituting the quantity $Z_bZ_c$ by $\beta Z_bZ_c$ where the correction factor $\beta$  assumes the values 0, 0.2, 1 and 2. 
It is seen 
that the smaller $\beta$ is, the less smooth is the joining of two parts of the curves of 
$\Delta_0(E)$ corresponding to $E>0$ and $E<0$ at $E=0$. Naturally, the effectiveness of the polynomial approximation of the function $\Delta_0(E)$ 
also drops with decreasing $\beta$. At $\beta=0$ (the Coulomb interaction is switched off) the  $\Delta_0(E)$ function turns into the ERF 
$k\cot\delta$ and ceases to coincide with the denominator of the amplitude $\tilde f_0(E)$ at $E<0$. 

The results obtained in this section corroborate and elucidate the conclusion drawn from the results of Sections III and IV, namely, that the polynomial approximation of the $\Delta$ function is more effective for heavier nuclear systems with larger values of the Coulomb parameter $\eta$.

\section{Conclusions}

In the present paper, within an exactly solvable model, we have investigated the applicability of the effective range function (ERF) and the 
$\Delta$ function suggested in Ref.~\cite{Sparen} for continuation of scattering data to the negative-energy region in order to determine ANCs. The 
$d+\alpha$ and $\alpha+^{12}$C systems have been considered.
It is demonstrated that, if the system under consideration features two bound states with the same quantum numbers,
then the ERF and $\Delta$ functions have two poles: one in the positive-energy region and the other in the negative-energy region, between the energies corresponding to the two bound states. 
It is also shown that, 
if the system has more than one bound state with the same quantum numbers, then the method of the continuation in energy of the ERF or $\Delta$ functions practically allows one to determine the binding energy and the ANC for the highest state only. To determine the features of other (lower-lying) bound states, one should apply alternative methods, e.g., the method of analytic continuation of  differential cross sections of transfer reactions to the pole in the scattering angle or find peripheral transfer reactions populating the bound states of interest.

It is demonstrated that the approximation of the $\Delta$ function by the first several terms of its Taylor expansion can be successfully used to determine binding energies and ANCs for the nuclear systems with sufficiently large $Z$. The procedure is less effective for the systems with small 
$Z$. The criterion for the applicability of such an approximation is derived. 

The renormalized Coulomb-nuclear amplitude $\tilde f_l(E)$ was introduced in Ref.~\cite{Hamilton}. It was shown 
that the analytic properties of $\tilde f_l(E)$ on the physical sheet are similar to those of the scattering amplitude generated by the short-range potential. On the other hand, it was also stated \cite{Hamilton} that $\tilde f_l(E)$ possesses the essential singularity at $E=0$. These two assertions contradict each other since the scattering amplitude for the short-range potential does not possess an essential singularity at $E=0$. It is known that an arbitrary function  
$\varphi (z)$ has no definite limit at $z\to z_0$ if $z_0$ is a point of an essential singularity. In the vicinity of the essential singularity the function may take any value. The calculations performed within the model used in the present paper have shown that the amplitude $\tilde f_0(E)$ has a definite limit at $E\to 0$ that does not depend on the direction from which $E$ approaches zero. It means that the point $E=0$ is not an essential singularity point of $\tilde f_0(E)$. The amplitude $\tilde f_0(E)$ possesses the unitary cut $0\le E<\infty$ on which 
$\mathrm{Im}[\tilde f_0(E)]$ has a discontinuity. 

In the present paper, the approximate versions of the ERF and $\Delta$ functions have been constructed on the basis of Taylor expansions at  zero energy. Of course, there are alternative ways to construct the  approximate forms of these functions, e.g., by rational functions in the form of Pad\'e approximants. We expect that using Pad\'e approximants should not change the qualitative conclusions made above. The test calculations using Pad\'e approximants did not improve appreciably the unsatisfactory results obtained in Section III b for the polynomial approximation of the $\Delta$ function. 
Furthermore, though all calculations were performed for $l = 0$, we believe the inferences made in the present paper should be valid for arbitrary $l$.

\section*{ Acknowledgments}

This work was supported by the Russian Science Foundation Grant No. 16-12-10048 (L.D.B.) and 
the Russian Foundation for Basic Research Grant No. 16-02-00049 (D.A.S.).  A.S.K. acknowledges a support from the Australian Research Council. A.M.M.  acknowledges support from the U.S. DOE Grant No. DE-FG02-93ER40773 and the U.S. NSF Grant No. PHY-1415656.

\section*{Appendix}

Consider in more detail the function $R(E)$ (see Eq.~(\ref{regfun})) discussed in Section 5. Using the asymptotic expansion of 
$\psi(z)$ at $|z|\to\infty$ \cite{Olver} and inserting $\eta=\alpha_1/\sqrt{2\mu E}$, one can write $h(\eta)$ in the form of Eq.~(\ref{h-asymptE}), where  the residual term $U_n(z)$ is subject to 
\begin{equation}\label{resid}
|U_n(z)| \le \frac{|B_{2n}|}{2n\cos^{2n+1}(\arg(z)/2)|z|^{2n}}, \ |\arg(z)| < \pi.
\end{equation}
For positive energies ($E>0$), $z= i\eta= i \alpha_1/\sqrt{2\mu E}$. Therefore, $\arg(z)=\pi/2$. Then, taking into account $\cos(\pi/4)=1/\sqrt{2}$ we can write 
\begin{equation}\label{resid1}
|U_n(E)| \le   \frac{\sqrt{2}|B_{2n}|  2^n}{2n} \left( \frac{2\mu E}{\alpha_1^2} \right)^n.
\end{equation}
For negative energies ($E<0$), $z= i\eta= \alpha_1/\sqrt{2\mu |E|}$. Therefore, $\arg(z)=0$. Then, using $\cos(0)=1$ we have
\begin{equation}\label{resid2}
|U_n(E)| \le  \frac{|B_{2n}|}{2n} \left( \frac{2\mu |E|}{\alpha_1^2} \right)^n. 
\end{equation}
If the series (\ref{h-asymptE}) were convergent, then at $n \to \infty$ $U_n(z)\to 0$. However, the series (\ref{h-asymptE}) is asymptotic and the residual term behaves differently. With increasing $n$, $|U_n(E)|$ decreases but beginning with some $n$ it starts to grow unrestrictedly. The corresponding value of $n$ depends on $E$. It is useless to increase this value of $n$ since at this value the partial sum of the series 
(\ref{h-asymptE}) is the best approximation of the exact value of $h(\eta)$. It is natural to set this value equal to the maximal value of $n$ at which the following condition holds
\begin{equation}\label{nmax}
\left| \frac{U_{n+1}(E)}{U_n(E)} \right| < 1. 
\end{equation}
Evaluation of the residual term allows one to evaluate $n$ by setting $U_n(E)$ equal to its maximal value. Such evaluation is very strict, nevertheless it makes finding the upper boundary for $n$ possible.  

For positive energies the condition (\ref{nmax}) takes the form
\begin{equation}\label{nmax_plus}
\frac{2n|B_{2n+2}|}{(n+1)|B_{2n}|} \left( \frac{2\mu E}{\alpha_1^2} \right) < 1.
\end{equation}
For negative energies Eq.~(\ref{nmax}) becomes
\begin{equation}\label{nmax_minus}
\frac{n|B_{2n+2}|}{(n+1)|B_{2n}|} \left( \frac{2\mu |E|}{\alpha_1^2} \right) < 1. 
\end{equation}
Condition (\ref{nmax_plus}) is more strict than (\ref{nmax_minus}). If  Eq.~(\ref{nmax_plus}) holds for some values of $n$  and $E>0$, then  condition (\ref{nmax_minus}) also holds for the same $n$ but for  $E'=-E<0$. Therefore, in what follows, we will use the more strict condition (\ref{nmax_plus}) to analyze specific systems.

The maximal value of $n$ at given $E$ and, vice versa, the maximal value of $E$ at given $n$, depend on the quantity $\alpha_2=
2\mu/\alpha_1^2$. The smaller $\alpha_2$ is, the larger is $n$ for given $E$ or the larger is $E$ for given $n$. This means that the smaller 
$\alpha_2$ is, the better the exact function $h(\eta)$ is approximated by the function $h_n(\eta)$ (see, Eq.~(\ref{hn}) which is the partial sum of the series 
(\ref{h-asymptE}).

For the $d+\alpha$ system $\alpha_2=7.53$ MeV$^{-1}$. Let us approximate the function $R(E)$ by the polinomial of the second power in $E$, that is, by the first three terms of the sum~(\ref{h-asymptE}).  In that case $n=4$. The maximal value of energy $E_n>0$, at which the condition 
(\ref{nmax_plus}) holds, is determined by the equation
\begin{equation}\label{E_n}
E_n = \frac{1}{\alpha_2} \frac{(n+1)|B_{2n}|}{2n |B_{2n+2}|}.
\end{equation}
At $n=4$ Eq. (\ref{E_n}) results in $E_4=0.036532$ MeV. Thus the energy interval, in which the employed approximation can satisfactorily describe the exact function $R(E)$, is extremely narrow and is not seen in Fig. \ref{pic3_1_2}.
At the same time, for the $\alpha+^{12}$C system $\alpha_2=0.0933$ MeV$^{-1}$ and $E_4=2.9460$ MeV. Therefore, the `favorable' energy interval is by two orders 
broader than for the $d+\alpha$ system which results in the successful polinomial approximation of $R(E)$ (see Fig. \ref{pic3_1_4}).

\end{document}